\documentclass[10pt,twocolumn,aps,pra,superscriptaddress,showpacs,tightenlines,pdflatex,longbibliography]{revtex4-2}

\usepackage{amsthm}
\usepackage{gensymb}
\usepackage{amsmath}            
\usepackage{amssymb}           
\usepackage{mathtools}
\usepackage{graphicx}           
\usepackage{xcolor}
\usepackage{braket}
\usepackage{etoolbox}
\usepackage{breqn}
\usepackage{epstopdf}
\usepackage{array}
\usepackage{diagbox}
\usepackage{slashbox}

\usepackage{xcolor}
\usepackage{bm}

\makeatletter
\let\cat@comma@active\@empty

\usepackage[colorlinks,citecolor=blue,urlcolor=blue,bookmarks=false,hypertexnames=true]{hyperref}

\newcommand{\comment}[1]{}

\theoremstyle{plain}

\newtheorem{lemma}{Result}

\newcolumntype{L}[1]{>{\raggedright\let\newline\\\arraybackslash\hspace{0pt}}m{#1}}
\newcolumntype{C}[1]{>{\centering\let\newline\\\arraybackslash\hspace{0pt}}m{#1}}
\newcolumntype{R}[1]{>{\raggedleft\let\newline\\\arraybackslash\hspace{0pt}}m{#1}}

\usepackage[normalem]{ulem}

\begin{document}
\title{Visually quantifying single-qubit quantum memory}

\author{Wan-Guan Chang}
\affiliation{Department of Physics and Center for Quantum Frontiers of Research \&
Technology (QFort), National Cheng Kung University, Tainan 701, Taiwan}

\author{Chia-Yi Ju}
\email{chiayiju@mail.nsysu.edu.tw}
\affiliation{Department of Physics, National Sun Yat-sen University, Kaohsiung 80424, Taiwan}
\affiliation{Center for Theoretical and Computational Physics, National Sun Yat-sen University, Kaohsiung 80424, Taiwan}

\author{Guang-Yin Chen}
\affiliation{Department of Physics, National Chung Hsing University, Taichung 402, Taiwan}

\author{Yueh-Nan Chen}
\email{yuehnan@mail.ncku.edu.tw}
\affiliation{Department of Physics and Center for Quantum Frontiers of Research \&
Technology (QFort), National Cheng Kung University, Tainan 701, Taiwan}

\author{Huan-Yu Ku}
\email{huan.yu@ntnu.edu.tw}
\affiliation{Department of Physics, National Taiwan Normal University, Taipei 11677, Taiwan}

\begin{abstract}

To store quantum information, quantum memory plays a central intermediate ingredient in a network. The minimal criterion for a reliable quantum memory is the maintenance of the entangled state, which can be described by the non-entanglement-breaking (non-EB) channel. In this work, we show that all single-qubit quantum memory can be quantified without trusting input state generation. In other words, we provide a semi-device-independent approach to quantify all single-qubit quantum memory. More specifically, we apply the concept of the two-qubit quantum steering ellipsoids to a single-qubit quantum channel and define the channel ellipsoids. An ellipsoid can be constructed by visualizing finite output states within the Bloch sphere. Since the Choi-Jamio{\l}kowski state of a channel can all be reconstructed from geometric data of the channel ellipsoid, a reliable quantum memory can be detected.
Finally, we visually quantify the single-qubit quantum memory by observing the volume of the channel ellipsoid. 

\end{abstract}

\maketitle

\section{Introduction}\label{sec:intro}
A quantum memory stores input quantum information and, at a later time, processes the retrieval of the information~\cite{Lvovsky2009,Bhaskar2020,ChenPRXQuantum2021,Li2021PRXQuantum,Ma2021}. To function as a network, the minimal criteria of a reliable quantum memory is preserving an entangled resource~\cite{Holevo2008,Horodecki2003,Ruskai2003}. In the language of the most general channel formalism, a reliable quantum memory can be described by a non-entanglement-breaking (non-EB) channel. Clearly, quantum information tasks based on entangled resources (i.e., entanglement-based quantum key distribution~\cite{Ekert91,Gisin2002,CurrsLorenzo2021}, quantum communication~\cite{Ursin2007,ArminPRXQuantum2021,Anwer2021noiserobust,Ho2022}, and quantum teleportation~\cite{Bennett93,Anderson2020,Kopszak2021multiportbased}) are not functional if there is an entanglement-breaking (EB) quantum memory in the network.

To verify a functional quantum memory, many independent protocols were proposed with different assumptions, for instance, the untrusted measurement apparatus~\cite{Pusey2015,Budroni_2019,Ku2021-1,Vieira2022temporal}, non-local game test~\cite{Xiao2021}, and coherence estimation~\cite{Simnacher2019}. Recently, Rosset \textit{et al.}~\cite{Rosset2018} investigated how to detect quantum memory with minimal assumptions (see also the continuous-variable version in Ref.~\cite{Paolo2023arXiv}) by expending the measurement-device-independent task into the temporal domain~\cite{Buscemi2012,Branciard2013,Rosset2020,Schmid2020typeindependent,Lipka2021PRXQuantum,Zhao2020}.
Importantly, the minimal assumptions on verification of quantum memory are two trusted state generation devices. One of them is an input of the quantum memory, while the other is used to create a quantum correlation for certifying reliable quantum memory. 

In this work, we geometrically quantify all single-qubit quantum memories by introducing the concept of channel ellipsoids, which are conceptually similar to quantum steering ellipsoids used to characterize the entanglement between qubit-qubit systems~\cite{Shi_2011,Jevtic2014,Milne_2014,Ku20182,Zhang2019QSE}. To be more specific, we define the channel ellipsoid of a single-qubit quantum memory as the set of all output states in the Bloch sphere. Although a channel ellipsoid is defined by all output states, the geometrical data is accessible with a finite number of inputs, which is useful for experimental studies on quantum memories. In this construction, there is no presumption on the input quantum states, in the sense that one can construct the channel ellipsoids of a single-qubit quantum memory in a semi-device-independent scenario~\cite{Gallego2010,Liang2011,Wang2019,Miklin2021universalscheme}. 

To quantify all single-qubit quantum memory, we first show that the geometric data from the channel ellipsoid can reconstruct the quantum channel in the Choi-Jamio{\l}kowski (CJ) representation. Therefore, all the properties of the single-qubit quantum memory can be visually characterized by the channel ellipsoid. Note that this feature does not hold in the quantum steering ellipsoid, which will be discussed later. In this way, we further give a clear physical interpretation of the volume of the channel ellipsoid in terms of the quantification of the quantum memory.
Moreover, we can fully distinguish the unwanted quantum memories by providing the geometrical representations of EB channels with channel ellipsoids. These representations provide a simple metric for quantifying a quantum memory in terms of its volume. Finally, we consider the depolarizing and amplitude damping channels as two concrete examples and test our results with using the IBM quantum simulator.

The structure of this paper is as the following description. We first briefly review the definitions of quantum steering ellipsoid and quantum memory in terms of non-EB channel and their properties in Sec.~\ref{sec:Background}. In Sec.~\ref{sec:Channel ellipsoids}, we will give the intuition of our main idea for visualizing the channel ellipsoid enlightened by the quantum steering ellipsoid \cite{Jevtic2014}. Then, in Sec.~\ref{sec:Volumeofthechannel}, we will introduce how to use the volume of the channel ellipsoid to quantify quantum memory by connecting the volume with the quantum memory robustness. By using the IBMQ simulation, we explicitly show the boundary of the non-entanglement breaking channel in Sec.~\ref{sec:IBM}. Finally, we summarize our results and provide our outlook in Sec.~\ref{sec:conclusion}.

\begin{figure}
	\centering
    \hspace*{-0.4cm} 
	\includegraphics[width=0.5\textwidth]{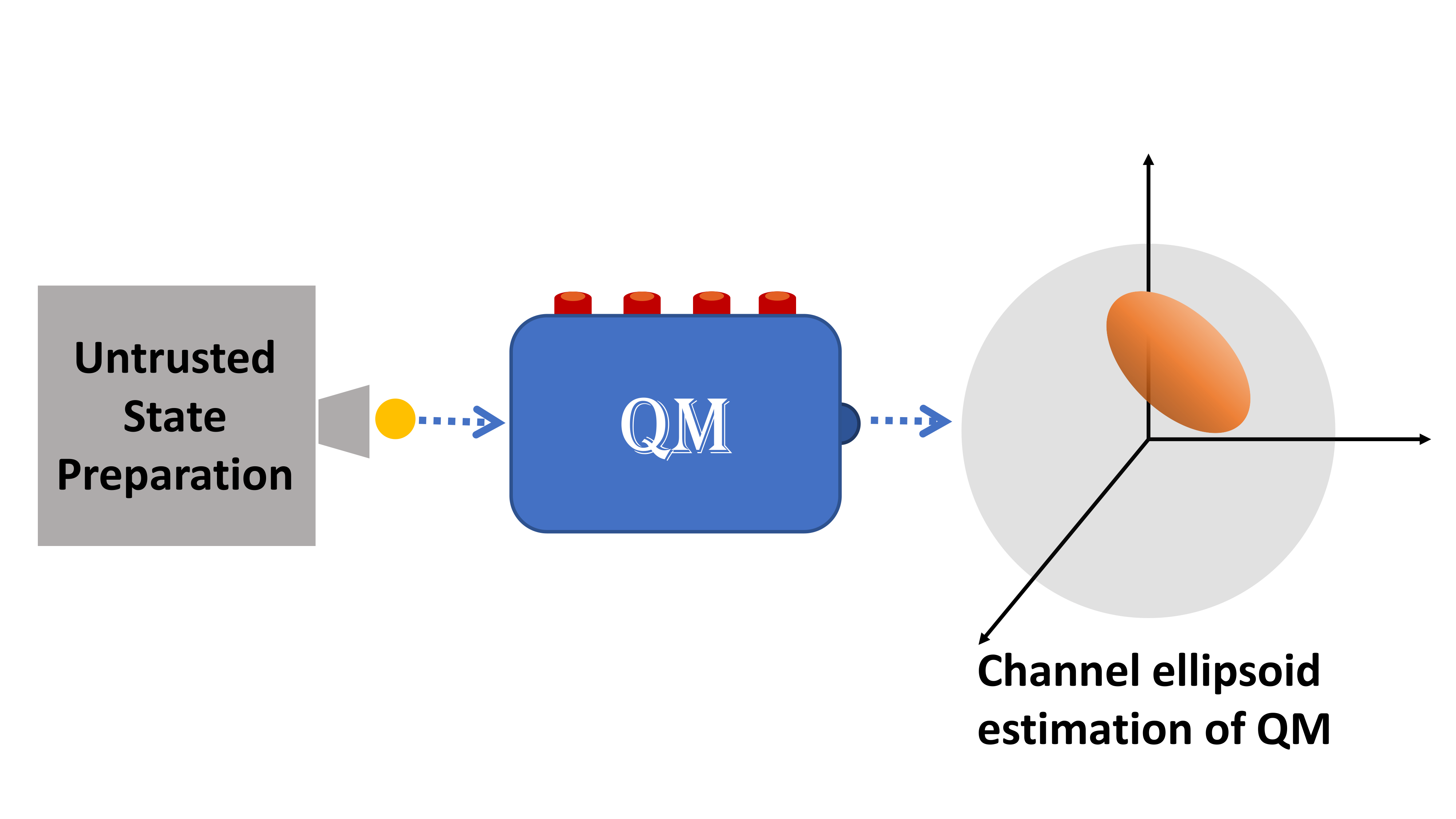}
	\caption{Schematic diagram of quantum memory. In this protocol, one can verify the quantum memory with channel ellipsoid based on semi-device-independent scenario.}
  \label{conceptfigure}
\end{figure}

\section{Background knowledge}\label{sec:Background}

\subsection{Quantum steering ellipsoid}\label{sec:QSE}
We briefly recall the concept of the quantum steering ellipsoid~\cite{Shi_2011,Jevtic2014,Milne_2014}, which is a visualization approach to study the two-qubit entangled state. Any two-qubit state shared by Alice and Bob can be expressed in the Pauli bases
$\rho^{\rm{AB}}=\openone/4+\sum_{i=1}^3 A_i\sigma_i\otimes\openone+\sum_{j=1}^3 B_j\openone\otimes\sigma_j+1/4\sum_{i,j=1}^3 T_{ij}\sigma_i\otimes\sigma_j$, where $\sigma_i$ denotes the Pauli matrices, $\mathbf{A}=(A_1,A_2,A_3)^{T}$ [$\mathbf{B}=(B_1,B_2,B_3)^{T}$], with $T$ standing for transpose, is the Bloch vector of Alice's (Bob's) reduced state, and $\mathbf{T}$ is a $3\times 3$ matrix representing the correlation of the state. 
Here, we in particular use the superscript $\rm{AB}$ to emphasize $\rho^{\rm{AB}}$ is a spatially separated state shared by Alice and Bob.
It gives rise to the quantum steering ellipsoid $\mathcal{E}^{\rm{B}}$ of Bob with center $\mathbf{C}^{\rm{B}}=\dfrac{\mathbf{B}-\mathbf{T}^{\text{T}}\mathbf{A}}{1-A^2}$ where $A^2 \equiv \mathbf{A}^{\text{T}} \mathbf{A}$. The orientation and semiaxes lengths $l_i=\sqrt{\lambda_i}$ can be determined by the eigenvectors and eigenvalues of the ellipsoid matrix
\begin{equation}
Q_{\rm{B}}=\frac{1}{1-A^2}(\mathbf{T}^{\text{T}}-\mathbf{B}\mathbf{A}^{\text{T}})\left(\openone+\frac{A A^{\text{T}}}{1-A^2}\right)(\mathbf{T}-\mathbf{A} \mathbf{B}^{\text{T}}).
\end{equation}

The operational definition of an ellipsoid $\mathcal{E}^{\rm{B}}$ represents the collection of all steered states generated by Alice's choices of measurements. The ellipsoidal surface is constituted when Alice considers all projectors in her subspace, while the most general quantum measurement, described by the positive operator-valued measure, generates the state inside the ellipsoid~\cite{Jevtic2014}. It has been shown that any quantum steering ellipsoids constructed by a separable state $\rho^{\rm{AB}}=\sum_ip(i)\rho^{\rm{A}}(i)\otimes\rho^{\rm{B}}(i)$ is always 
inside a tetrahedron which is inside a Bloch sphere. 
The definition of a quantum steering ellipsoids is reminiscent of the Einstein–Podolsky–Rosen steering~\cite{Wiseman2007,DCavalcanti16,UolaRev2020,XiangPRXQ2022}, in which the correlation contains not only Bob's steered states but also the probability distribution generated by Alice's measurements.
Due to the similar operational definitions, several approaches are proposed to connect quantum steering ellipsoid with Einstein–Podolsky–Rosen steering; e.g., very recently, it has been shown that Bob's steering ellipsoid can be used to predict the quantum steerability of a quantum state~\cite{McCloskey2017PRA,Wang2023arXiv,Ku2023arXiv}.
During a steering experiment, Alice's measurement may not be fully characterized, which can be applied to one-sided device-independent quantum information processing~\cite{Branciard2012,Skrzypczyk2018,Tan2021,Ku2022NC,Hsieh2023arXiv}.

\subsection{Quantum memory}\label{sec:Quantum memory}

Without loss of the generality, the effects of any quantum devices (e.g., quantum memory, quantum repeater, and quantum communication line) can be described by a quantum channel $\Lambda$, which is a complete-positive trace-preserving (CPTP) map.
The properties of a quantum channel can be characterized by a corresponding CJ state~\cite{Choi1975285,Jamiokowski1972}, defined as
\begin{equation}
\rho^\Lambda=\text{id}\otimes\Lambda(\ket{\Psi}\bra{\Psi}),
\label{Eq:CJstate}
\end{equation}
where $\text{id}$ is an identity channel and $\ket{\Psi}=\sum_i 1/\sqrt{N}\ket{i}\otimes\ket{i}$ is the maximally entangled state with local dimension $N$. 
Similar to the superscript in $\rho^{\rm{AB}}$, we use the superscript $\Lambda$ to emphasize $\rho^\Lambda$ is the CJ state of the channel $\Lambda$.
The output of the channel with an input state $\rho$ can be expressed as 
\begin{equation}
\Lambda(\rho)=\text{Tr}_{\rm{in}}\left( \openone\otimes\rho^{\text{T}} \rho^\Lambda\right),
\label{Eq:output_CJ}
\end{equation}
where $\text{T}$ denotes the transpose.
If a channel is CPTP, the reduced state of input space for any CJ state must be maximally mixed~\cite{Chiribella2009}, namely
\begin{equation}\label{Eq: marginal of CJ}
\text{Tr}_{\rm in}\left(\rho^\Lambda\right)=\openone/d_{\rm in},
\end{equation}
where $d_{\rm in}$ is the dimension of the input Hilbert space.

A perfect quantum memory should preserve the quantum information for a period of time without introducing any disturbance. In this sense, the perfect quantum memory can be described by an identity channel (or unitary channel). Recently, the minimal criterion of a quantum memory was proposed as the preservation of the entangled state~\cite{Rosset2018}. In the channel terminology, a functional quantum memory cannot be EB, namely,
\begin{equation}
    \Lambda_{\rm{EB}}(\rho)=\sum_k\text{Tr}[M_k\rho]\rho'_k,
\end{equation}
where $M_k$ denotes the generalized quantum measurement (positive operator-valued measure), and $\rho'_k$ is the quantum state conditioned on the measurement outcome $k$. The set of all EB channels is denoted as $\mathcal{EB}$. Because the entanglement is destroyed by the measurement, we can easily observe that the CJ state of an EB channel is a separable state~\cite{Horodecki2003,Holevo2008} $\rho^{\Lambda_{\rm{EB}}}=\sum_i p(i)\rho(i)\otimes\sigma(i)$, where $\rho(i)$ and $\sigma(i)$ are quantum states sharing with classical correlation $i$. One can use the memory robustness to measure a quantum memory $\Lambda$~\cite{Xiao2021}:
\begin{equation}
    QM(\Lambda)=\min_{\Lambda'}\left\{t\geq 0 \, \left| \,\frac{\Lambda+t\Lambda'}{1+t} \right. \in\mathcal{EB}\right\},
\label{Eq:memory robustness}
\end{equation}
where $\Lambda'$ can be seen as an arbitrary channel with noisy parameter $t$ added on the given quantum memory $\Lambda$. This quantity can be easily computed via a semidefinite program~\cite{boyd_vandenberghe_2004} when considering the CJ representation. Note that the memory robustness is a proper measure under the resource theory of quantum memory; i.e., it is a nonincreasing function under any quantum-memory free operation~\cite{Rosset2018}, which maps any EB channel to the other EB channel.

\section{Channel ellipsoids}\label{sec:Channel ellipsoids}

Due to the one-to-one mapping between quantum channel and quantum state [cf., Eq.~\eqref{Eq:CJstate}], we can use quantum steering ellipsoids~\cite{Jevtic2014} to study the properties of a single-qubit quantum channel, visually. More specifically, we can span the CJ state of any single-qubit channel in the Pauli basis, namely
\begin{equation}
\rho^\Lambda=\openone/4+\sum_{i=1}^3 A_i\sigma_i\otimes\openone+\sum_{j=1}^3 B_j\openone\otimes\sigma_j+1/4\sum_{i,j=1}^3 \Theta_{ij}\sigma_i\otimes\sigma_j.
\label{Eq:CJstate in Pauli basis}
\end{equation}
Due to the reduced-state condition of the CJ state [cf., Eq.~\eqref{Eq: marginal of CJ}], $\mathbf{A}$ must be a null vector. Similar to the case in quantum steering ellipsoid, $\mathbf{B}=(B_1,B_2,B_3)^{\text{T}}$ represents the Bloch vector of the output state with input being the maximally mixed state. Finally, $\mathbf{\Theta}$, a square matrix with components being $\Theta_{ij}$, is used to describe the correlations between the input and output states. 
Thus, the set of all output states can be visualized by the channel ellipsoid with the center $\mathbf{C}^{\Lambda}=(B_1,B_2,B_3)^{\text{T}}$. The orientation and semiaxes lengths $l_i=\sqrt{\lambda_i}$ of a channel ellipsoid can be determined by the eigenvectors and eigenvalues $\lambda_i$ of the ellipsoid matrix 
\begin{equation}
Q^{\Lambda}=\mathbf{\Theta}^{\text{T}}\mathbf{\Theta}.
\label{Eq:ellipsoid matrix}
\end{equation}

The operational definition of a channel ellipsoid is the collection of all output states of the channel. Since we only focus on the output states, the input state generations can be seen as a black box. In this sense, the construction of a channel ellipsoid is independent of the input quantum states. In other words, the channel ellipsoids can be constructed with only trusting measurement apparatus at output space, which leads to the semi-device-independent scenario.

Furthermore, from the geometric data $(\mathbf{C}^\Lambda,Q^{\Lambda})$, one can faithfully reconstruct the CJ state of a channel $\rho^\Lambda$ because $\mathbf{A}$ is a null vector.  
In this way, we can use the standard approaches (i.e., Peres–Horodecki criterion~\cite{Peres1996,HORODECKI19961}, and entanglement witness~\cite{Ghne2009}) to probe the non-EB channels with only the output information. 
Therefore, all the single-qubit quantum memory can be certified in a semi-device-independent scenario. It is worth mentioning that because the CJ state of every EB channel is separable~\cite{Ruskai2003,Horodecki2003}, any corresponding channel ellipsoid can be enclosed by a tetrahedron~\cite{Wootters1998,Jevtic2014}.
This property is very different from the case of the quantum steering ellipsoids in which the geometric data of an ellipsoid from only one side is not sufficient to rebuild the density operator $\rho^{\rm AB}$ of a two-qubit entangled state~\cite{Jevtic2014}.

Finally, although the definition of the channel ellipsoid is the set of all output states, the channel ellipsoid can be constructed with a finite number of the output states. It is because to obtain a unique solution of an ellipsoid function, nine inputs (or, equivalently nine output quantum states) are sufficient, which leads to an experimentally feasible test of a single-qubit quantum memory in a channel ellipsoid manner (see also Sec.~\ref{sec:IBM}).

\section{Quantifying quantum memory with the volume}\label{sec:Volumeofthechannel}
As we mentioned in the previous section, we can consider the geometric data $(\mathbf{C}^\Lambda,Q^{\Lambda})$ of a channel ellipsoid to test a single-qubit quantum memory. We now discuss the properties of the volume of the channel ellipsoid, determined by 
\begin{equation}
V^\Lambda=\frac{4\pi}{3}|l_1l_2l_3|.
\end{equation}
Later, we show that the volume of the channel ellipsoid provides an upper bound on the memory robustness. Therefore, the larger volume of the channel ellipsoid implies better functional quantum memory. 

Before we show our main result in this section, we first recall the concurrence and the negativity of a bipartite quantum state $\rho^{\rm AB}$, shared by Alice and Bob. The concurrence is used to estimate the number of the maximally entangled states required to construct the given state $\rho^{\rm AB}$ by local operations with classical communication~\cite{Wootters1998}. The concurrence can be computed by 
\begin{equation}
E_{C}(\rho^{\rm AB})= \text{max}\{0,\lambda_1-\lambda_2-\lambda_3-\lambda_4\},
\end{equation}
where $\lambda_i$ (in the decreasing order) are the eigenvalues of $\sqrt{\sqrt{\rho^{\rm AB}}\hat{\rho}^{\rm AB}\sqrt{\rho^{\rm AB}}}$ with $\hat{\rho}^{\rm AB}=(\sigma_2\otimes\sigma_2)\rho^{\rm{AB}\dagger}(\sigma_2\otimes\sigma_2)$. Here, $\dagger$ denotes the conjugate transpose. It has been shown that the concurrence is a lower bound on the volume of the quantum steering ellipsoids~\cite{Milne_2014}. 

On the other hand, the negativity is another entangled measure, based on the Peres–Horodecki criterion~\cite{Peres1996,HORODECKI19961}. In short, since the partial transposition on any qubit-qubit entangled states does not preserve the positivity, there must exist, at least, one negative eigenvalue for the state after partial transposition. The negativity is defined as the absolute sum of the negative eigenvalues of the state after the partial transpose, namely 
\begin{equation}
E_{N}(\rho^{\rm AB})=\sum_{\lambda_i<0}\| \lambda_i\|,
\end{equation}
where $\lambda_i$ are the eigenvalues of the partial-transposed state.

Both concurrence and negativity are entanglement measures in the sense that they are monotonically decreasing functions under any local operations with classical communications. However, they may not be proper quantum memory measures because (1) the free operations under the frameworks of resource theory of entanglement and memory are different, and (2) the mathematical transformations of a state and a channel under their corresponding free operations are different. 
For instance, it has been shown that the negativity of the CJ state of a quantum channel is the quantum memory measure under the particular quantum-memory free operation~\cite{Ku2021-1}. Note, the memory robustness cannot be increased under ``any" memory free operation.

With the concurrence and negativity, we can now introduce:
\begin{lemma}
The volume of the channel ellipsoid is an upper bound on the memory robustness:
\begin{equation}
    QM(\Lambda)\leq \left(\frac{3V^{\Lambda}}{4\pi}\right)^{\frac{1}{4}}.
\label{result1}
\end{equation}
\end{lemma}
\emph{Proof.---} First, we recall the fact that, for any state $\rho^{\rm AB}$, the volume of the ellipsoid with respect to Bob is bounded by~\cite{Milne_2014}
\begin{equation}\label{Eq: concurrence and volume}
    E_C(\rho^{\rm AB})\leq \mathcal{R}_{\rm A}^{-1}\left(\frac{3V^{B}}{4\pi}\right)^{\frac{1}{4}},
\end{equation}
where $\mathcal{R}_{\rm A}^{-1}=1/\sqrt{1-A^2}$ is the Lorentz factor. We note that if we replace Bob's ellipsoid with Alice's one by swapping $\mathbf{A} \leftrightarrow \mathbf{B}$ and $\Theta \leftrightarrow \Theta^{\text T}$, the above-bounded relation also holds. From Refs.~\cite{Verstraete2001} and \cite{Frank2003}, we have $E_C(\rho^{\rm AB})\geq E_N(\rho^{\rm AB})$ and $E_N(\rho^{\rm AB})\geq E_R(\rho^{\rm AB})$, respectively. Here, $E_R(\rho^{\rm AB})$ is the entanglement robustness defined in the same way in Eq.~\eqref{Eq:memory robustness}, but substituted the channel and EB channel with a bipartite quantum state and a separable state.

We now apply the property $\mathbf{A}=0$ of the CJ state on the Lorentz factor leading $\mathcal{R}_{\rm A}^{-1}=1$. Since the CJ states can be seen as a subset of the entangled state, we can substitute the entangled state in $E_C(\rho^{\rm AB})$, $E_N(\rho^{\rm AB})$, and $E_R(\rho^{\rm AB})$ with the CJ state. With the transitive relation, we complete the proof. $\square$

Despite the fact that it is well-established that all the outputs of the channel form an ellipsoid~\cite{NielsenBook}, the physical meaning of the volume of the channel ellipsoid was, to some extend, unclear. Our result provides an explicitly quantitative relationship between the volume and the quantum memory. More specifically, although the relation between the volume of a channel ellipsoid and the concurrence of the CJ state of the channel can be derived [cf., Eq.~\eqref{Eq: concurrence and volume}], the quantum-memory monotonicity of the concurrence of a CJ state remains unclear. Here, we provide the physical meaning of the volume of the channel ellipsoid by instead showing the relationship with the quantum memory robustness, which is a valid quantum memory measure. In this sense, we call the LHS of Eq.~\eqref{result1} is volume estimated upper bound of memory robustness. Moreover, since the nature of the CJ states [cf., Eq.~\eqref{Eq:CJstate in Pauli basis}] is equivalent to the canonical form of the qubit-qubit entangled states, all the results in Ref.~\cite{Milne_2014} are also valid in the channel scenario. For instance, the maximal volume of channel ellipsoid with the associated center distance $|\mathbf{C}^{\Lambda}|$ is determined by the channel ellipsoid of the amplitude damping channel~\cite{Milne_2014}.

\section{Implementation with IBM quantum device}
\label{sec:IBM}
In this section, we demonstrate the channel ellipsoids of depolarizing channels and amplitude damping channels via IBMQ simulator. We use a depolarizing (amplitude damping) channel to show how to certify (quantify) quantum memories with the channel ellipsoids by the tetrahedron (volume) condition. 

\subsection{Depolarizing channel}
The depolarizing channel is widely used in quantum information to apply the quantum correlation certification~\cite{Heinosaari2015,RodriguezPRXQuantum2021,EtxezarretaMartinez2021,Girard2021} and measurement mitigation~\cite{Maciejewski2020mitigationofreadout,Zhang2020,Cardani2021,SuzukiPRXQuantum2022}.
The depolarizing channel is a convex combination of the input quantum state $\rho$ and the maximally mixed state with the mixing parameter $P$, namely,
\begin{align}
    \mathcal{E}_{D} ( \rho ) = \frac{1-P}{2}\openone + (P) \rho \, . \label{depolarizingchannel}
\end{align}
It has been shown that the depolarizing channel is EB when $P\geq \frac{1}{3}$.

\begin{figure}
	\centering
	\includegraphics[width=0.5\textwidth]{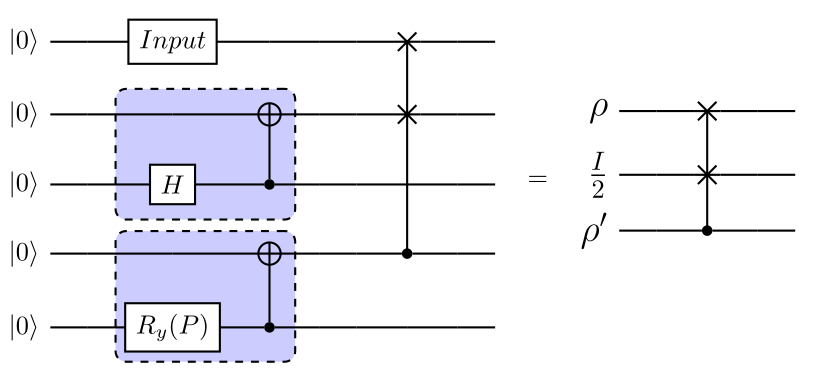}
	\caption{Circuit implementation of the depolarizing channel. The input states are prepared by scanning uniformly generated unitary operations on the Bloch sphere. The second and third qubits simulate the maximally mixed state by applying the Hadamard gate and CNOT gate on the ancilla. Therefore the reduced state of the second qubit becomes the maximally mixed state. Finally, the fourth and fifth qubits are used to manipulate the parameter $P$ of state $\rho' = (1-P) \ket{0} \bra{0} + P \ket{1} \bra{1}$ which demonstrates probability in Eq~\eqref{depolarizingchannel}. Finally, the Fredkin gate is applied on the first, second, and fourth qubits such that the output state can be seen as the input state after the depolarizing channel.}
  \label{DPchannel}
\end{figure}

Here, we use five qubits to demonstrate the ellipsoid of a depolarizing channel by the circuit model (see also Fig.~\ref{DPchannel}).
We first construct the input states $\rho$ uniformly on the Bloch sphere by inserting uniform unitary $U(\theta,\psi)$ on the initial state $\ket{0}$. Since the depolarizing channel is a convex combination of the input state and the maximally mixed state, we need to generate a maximally mixed state. Here, we use the property that the marginal of the maximally entangled state is a maximally mixed state to finish the construction. Now, we use the Fredkin gate with the control qubit to be the classical mixture of the $\ket{0}$ and $\ket{1}$. This classical mixture can be achieved by considering, again, the reduced state of the pure entangled state. Finally, the control qubit can flip the input state into the maximally mixed state with probability $1-P$ (see the circuit of the depolarizing channel in our composer is shown as Fig.~\ref{DPchannel}). We note that the probability $P$ here in IBMQ composer can be computed by $P = 1-\sin^2 \frac{\theta}{2}$.

\begin{figure}[h!]
	\centering
	\includegraphics[width=0.5\textwidth]{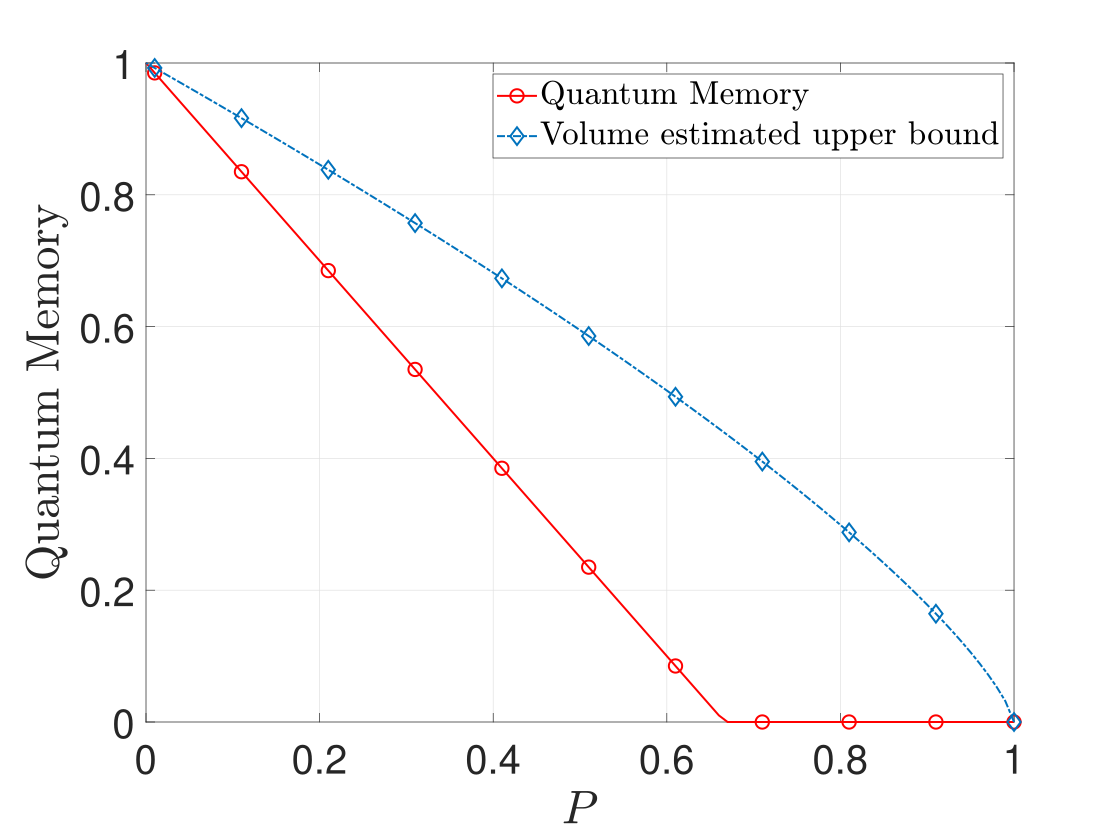}
	\caption{Memory robustness (red circles and solid curve) and volume estimated upper bound (blue diamonds and dashed curve) of the depolarizing channel, characterized by the mixing parameter in Eq.~\eqref{depolarizingchannel}. When the mixing parameter decreases, both the memory robustness and volume estimated upper bound monotonically decrease too. After $P\geq \frac{2}{3}$, the channel is EB leading to the vanishing memory robustness. }
 \label{Fig:robusntess of dep}
\end{figure}

Since the above circuit required five qubits and multiple CNOT gates, we implemented this circuit on the IBM quantum simulator. The simulation results of channel ellipsoids from IBM quantum simulator are presented in Table.~\ref{ellipsoidtable_depolarizing}. We also present the comparison of the memory robustness and the volume estimated upper bound in Fig.~\ref{Fig:robusntess of dep}. From the results, one can see that the ellipsoids shrank when the mixing parameter $P$ decreases (see also Table.~\ref{ellipsoidtable_depolarizing}). One can also observe that if the channel ellipsoids of the depolarising channels are in the range of $P\leq \frac{1}{3}$, the channels cannot be used as a functional quantum memory. This is because the ellipsoids cannot violate the separability constraint i.g., the ellipsoids are inside the tetrahedrons in the Bloch sphere.

\begin{table*}[t!]
    \centering
    \begin{tabular}{|C{2.7cm} C{2.7cm} C{2.7cm} C{2.7cm} C{2.7cm} C{2.7cm}|}
        \hline
        $\theta = 0$ & $\theta = \frac{\pi}{4}$ & $\theta = \frac{2\pi}{4}$ & $\theta = 0.609\pi$ & $\theta = \frac{3\pi}{4}$ & $\theta = \pi$\\
        \hline
        \includegraphics[width=0.16\textwidth]{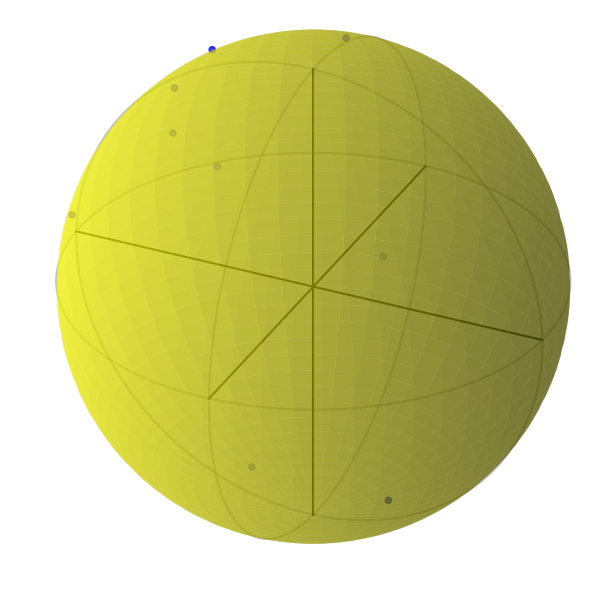}
        &
        \includegraphics[width=0.16\textwidth]{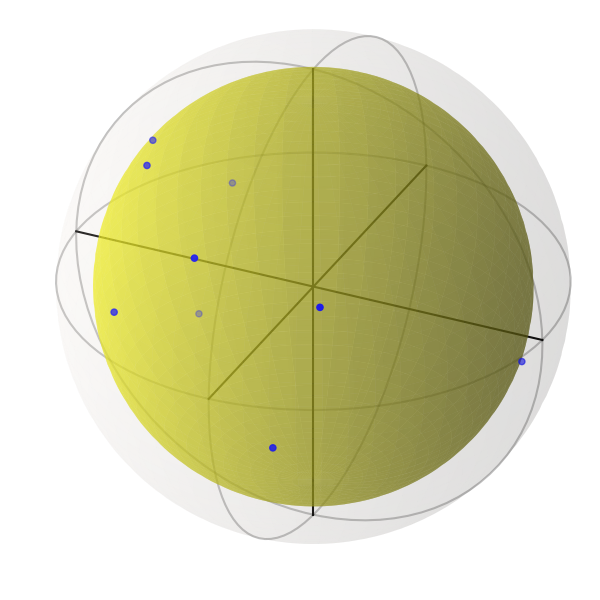}
        & 
        \includegraphics[width=0.16\textwidth]{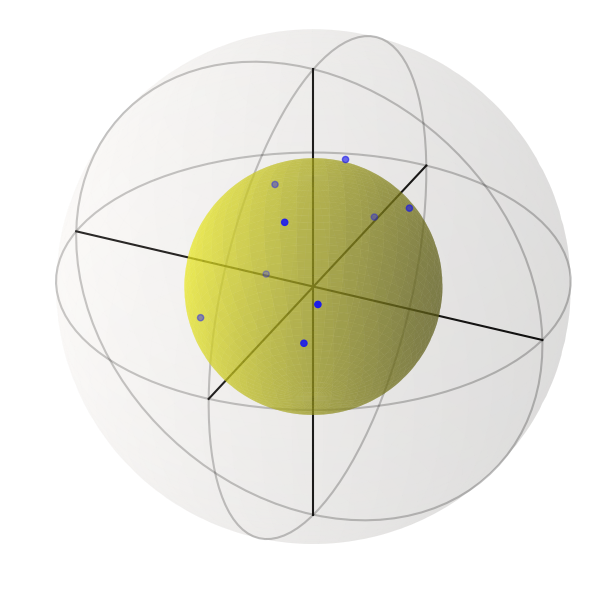}
        &
        \includegraphics[width=0.16\textwidth]{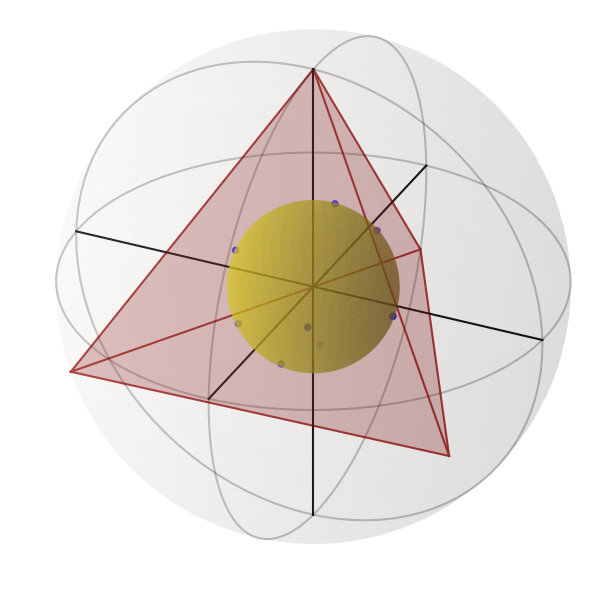}
        &
        \includegraphics[width=0.16\textwidth]{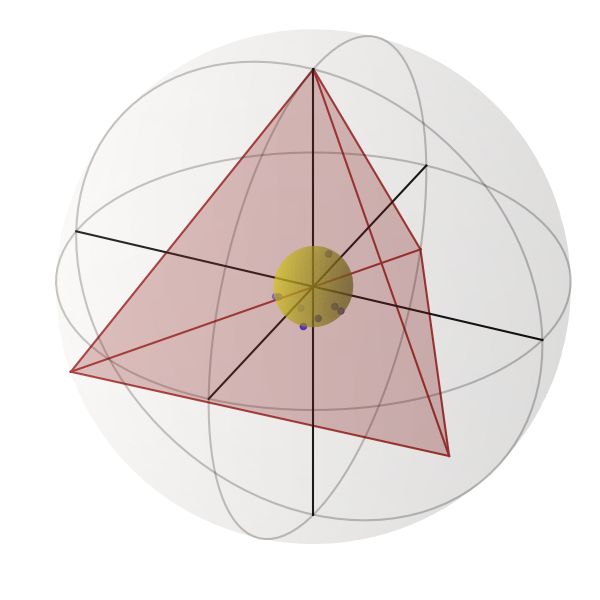}
        &
        \includegraphics[width=0.16\textwidth]{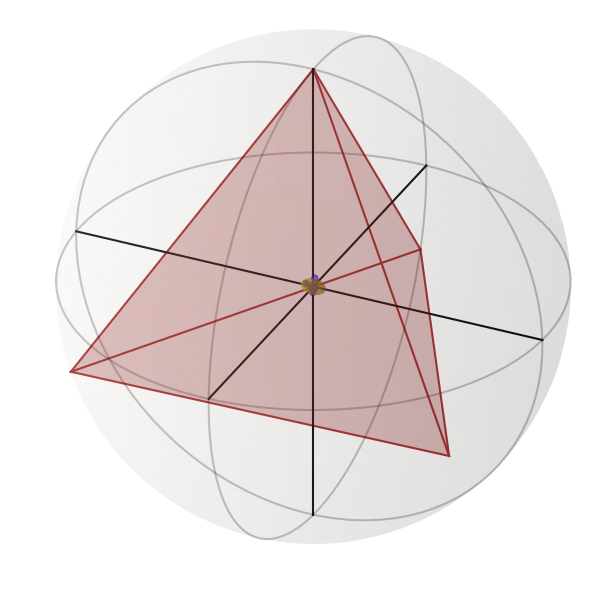}\\
        \hline
        \end{tabular}
    \caption{Volume of depolarizing channel's ellipsoid. Here, $P=1-\sin^2 \frac{\theta}{2}$ is the mixing parameter. When $\theta$ increases, $P$ and the volumes of channel ellipsoids simultaneously decrease too. When $\theta=0.609\pi$, which corresponds to $P=0.6679$, there exists the tetrahedron in the Bloch sphere such that the ellipsoid is inside the tetrahedron.}
    \label{ellipsoidtable_depolarizing}
\end{table*}

\subsection{Amplitude damping channel}
\begin{figure}[h!]
	\centering
	\includegraphics[width=0.45\textwidth]{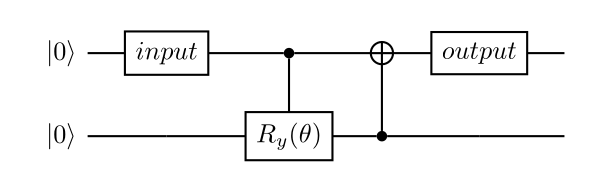}
	\caption{Circuit implementation of the amplitude damping channel. The input here is also the prepared state by scanning uniformly the Bloch sphere. The rest of the construction is a typical amplitude damping channel by applying the control $R_y$ gate to simulate the energy dissipation.\label{apchannel}}
\end{figure}

The amplitude damping channel is commonly used to characterize the dissipation of energy with the dissipation rate $\gamma$. The evolution of the input state after the amplitude damping channel can be described as 
\begin{equation}
\begin{split}
    & (a\ket{0}_{\text{in}} + b\ket{1}_{\text{in}})\ket{0}_{\text{anc}} \rightarrow \\
    & (a\ket{0}_{\text{out}} + b \sqrt{1-\gamma} \ket{1}_{\text{out}})\ket{0}_{\text{anc}} + \sqrt{\gamma}\ket{0}_{\text{out}}\ket{1}_{\text{anc}},
\end{split}
\end{equation}
where the subscript `in' and `out' are the input and output of the circuit and subscript 'anc' is the ancilla qubit, or equivalently the environment system (see Fig.~\ref{apchannel}). If environment goes to $\ket{1}_{\text{anc}}$, the output state will become $\ket{0}_{\text{out}}$. Otherwise, the output state will be prepared as the combination of $\ket{0},\ket{1}$. 
Under this construction, the Kraus representation of the amplitude damping channel is characterized by 
\begin{align}
    \mathcal{E}_{AD}( \rho ) = E_0 \rho E_0^\dagger + E_1 \rho E_1^\dagger ,
\end{align}
where 
\begin{align}
    E_0 & = 
    \begin{pmatrix}
        1 & 0 \\
        0 & \sqrt{1-\gamma}
    \end{pmatrix} \nonumber\\
    E_1 & = 
    \begin{pmatrix}
        0 & \sqrt{\gamma} \\
        0 & 0
    \end{pmatrix}.
    \label{Eq: Kraus of amp}
\end{align}
Note that the amplitude damping channel is not an EB channel.

To demonstrate the amplitude damping channel in the circuit model, we use a similar approach to the case in the depolarizing channel to construct the input state. After the initialization, we apply a control-$Ry$ gate which rotates the target qubit from $\ket{0}$ to an angle $\theta$ along the y-axis. 
With this construction, the dissipation rate satisfies $\gamma=\sin^2(\theta)$.
Then, a CNOT gate is implemented to finish the amplitude damping channel (see Fig.~\ref{apchannel}). 

\begin{figure}[h!]
	\centering
	\includegraphics[width=0.5\textwidth]{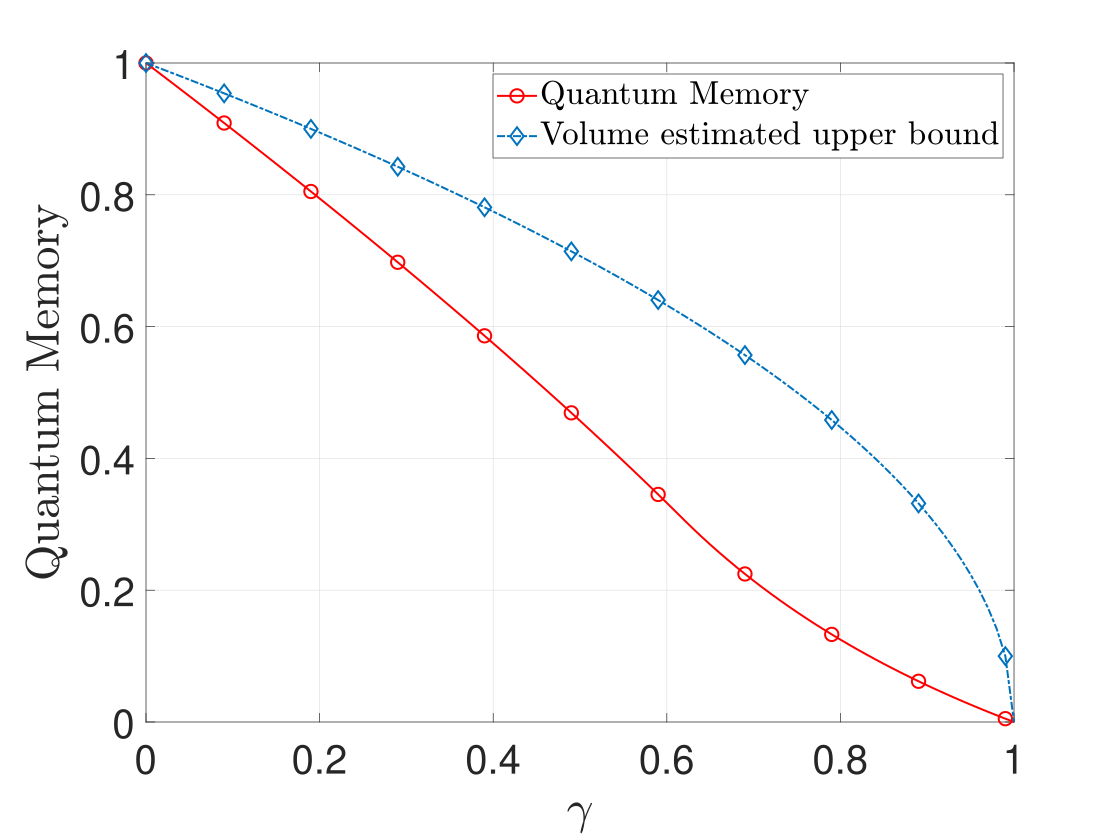}
	\caption{Memory robustness (red circles and solid curve) and volume estimated upper bound (blue diamonds and dashed curve) of the amplitude damping channel, characterized by the dissipation rate in Eq.~\eqref{Eq: Kraus of amp}. When the dissipation rate increases, both the memory robustness and volume estimated upper bound monotonically decrease too. Since the amplitude damping channel is not EB unless $\gamma=1$, both the memory robustness and the volume estimate upper bound are always non-zero. }
 \label{Fig:robusntess of amp}
\end{figure}

The demonstration results by IBM quantum simulation with $P\in \{0,\frac{\pi}{4},...,\pi\}$ are shown in Table.~\ref{ellipsoidtable_amplitude}. From the data points, one can compute the volume of the ellipsoid (see also Fig.~\ref{Fig:robusntess of amp}). As expected, both the memory robustness and volume estimated upper bound monotonically decrease as the dissipation rate increases. Note that both the memory robustness and the volume estimate upper bound are non-zero because the amplitude damping channels are not EB channels. 
Our result also shows that the amplitude damping channel provides the maximal volume of the channel ellipsoids with the associated center distance $|\mathbf{C}^{\Lambda}|$~\cite{Milne_2014}.

\begin{table*}[t!]
    \centering
    \begin{tabular}{|C{2.7cm} C{2.7cm} C{2.7cm} C{2.7cm} C{2.7cm}|}
        \hline
        $\theta = 0$ & $\theta = \frac{\pi}{4}$ & $\theta = \frac{2\pi}{4}$ & $\theta = \frac{3\pi}{4}$ & $\theta = \pi$ \\
        \hline
        \includegraphics[width=0.16\textwidth]{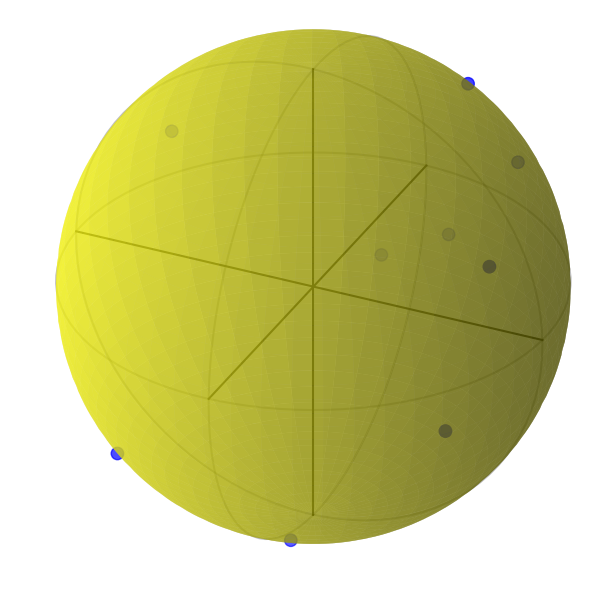}
        &
        \includegraphics[width=0.16\textwidth]{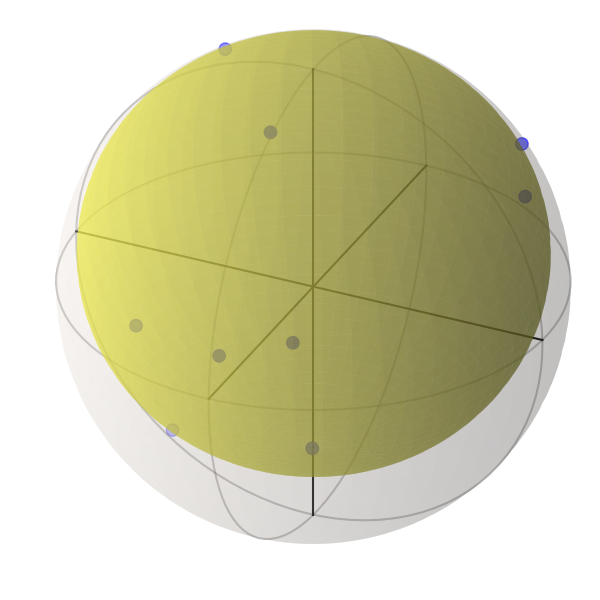}
        &
        \includegraphics[width=0.16\textwidth]{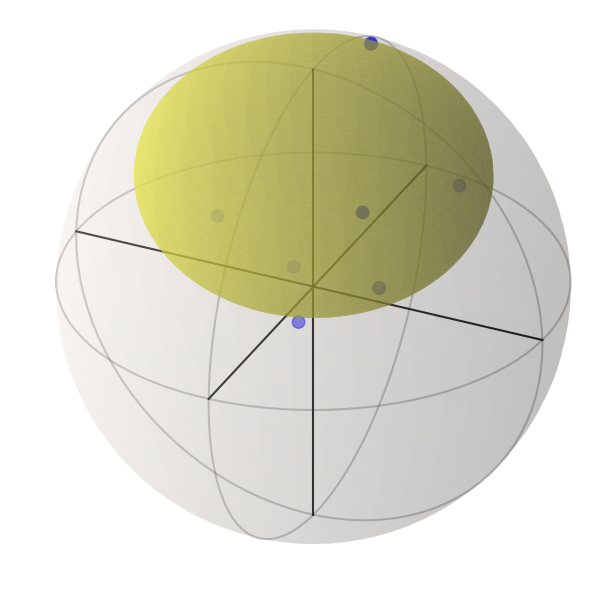}
        &
        \includegraphics[width=0.16\textwidth]{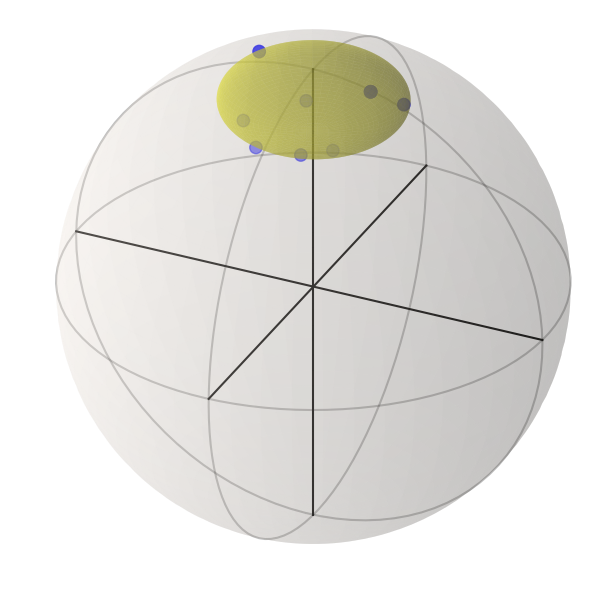}
        &
        \includegraphics[width=0.16\textwidth]{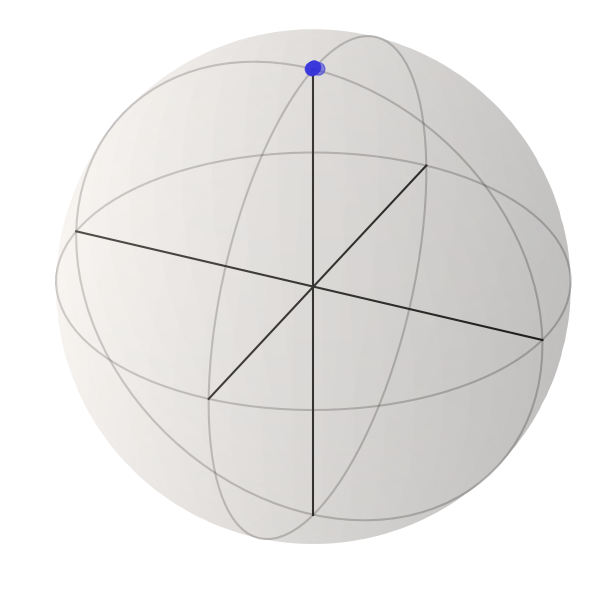} \\
        \hline
        \end{tabular}
    \caption{Volume of amplitude damping channel's ellipsoid. Here, we rotate the target qubit along the y-axis by angles $\{0, \pi/4, \pi/2, 3\pi/4, \pi \}$. When $\theta$ increases, the volumes of channel ellipsoids simultaneously decrease.}
    \label{ellipsoidtable_amplitude}
\end{table*}

\section{Discussion}\label{sec:conclusion}

In this work, we investigate the relationship between channel ellipsoids and quantum memories. We show that all functional single-qubit quantum memory can not only be certified with finite geometrical data of the ellipsoid but also be visually quantified with the volume of the ellipsoid. Therefore, we provide a clear operational definition of the volume of the ellipsoid. We also simulate our results with the memories described by depolarizing and amplitudedamping channels.

This work also raises many interesting questions. As we have shown that for a qubit channel, it is sufficient to construct the ellipsoid with a finite amount of output geometrical data. Can this property generalize to a high-dimensional channel? In other words, can we construct the Choi state of a high-dimensional channel with finite output states? In this work, we only show that the volume of a channel ellipsoid is an upper bound of the memory robustness.
Can we use the properties of an ellipsoid to define a quantum memory monotone?

\section*{Acknowledgments}
H.-Y.~K. is supported by the
Ministry of Science and Technology, Taiwan, (Grants No.~MOST 112-2112-M-003-020-MY3), and Higher Education Sprout Project of National Taiwan Normal
University (NTNU) and the Ministry of Education (MOE) in Taiwan.
This work is supported partially by the National Center for Theoretical
Sciences and Ministry of Science and Technology, Taiwan, Grants
No. MOST 110-2811-
M-006-546, and the Army Research Office (under Grant No.
W911NF-19-1-0081). 
C.Y.J. is partially supported by the National Science and Technology Council (NSTC) through Grant No. NSTC 112-2112-M-110-013-MY3. G.Y.C. is partially supported by the NCTS and NSTC through Grant No. 112-2112-M-005-006 and 112-2123-M-006 -001.
\appendix

%

\end{document}